\documentclass[%
 aip,
amsmath,amssymb,
reprint,%
nofootinbib
]{revtex4-1}
\usepackage[utf8]{inputenc}
\usepackage[T1]{fontenc}
\usepackage{mathptmx}
\usepackage{etoolbox}
\usepackage{lipsum}
\usepackage{graphicx}
\usepackage{amssymb}
\usepackage{float}
\usepackage{url}
\usepackage{soul}
\usepackage[labelformat=simple]{subcaption}

\usepackage{float}
\usepackage{calligra}
\usepackage{calrsfs}
\usepackage{xcolor}
\usepackage{enumitem}
\usepackage{graphicx}
\usepackage{dcolumn}
\usepackage{bm}
\usepackage{tabularx,ragged2e,booktabs,caption}
\usepackage{xcolor}
\usepackage[left=2cm,right=2cm,top=2cm,bottom=2cm]{geometry}

\definecolor{LWcolor}{RGB}{0,0,255} 

\newcommand{\mycomment}[1]{}
\usepackage{icomma}

\setcitestyle{numbers,square}

\definecolor{HGcolor}{RGB}{255,20,147} 

\definecolor{GScolor}{RGB}{0,0,0} 

\makeatletter
\def\@email#1#2{%
 \endgroup
 \patchcmd{\titleblock@produce}
  {\frontmatter@RRAPformat}
  {\frontmatter@RRAPformat{\produce@RRAP{*#1\href{mailto:#2}{#2}}}\frontmatter@RRAPformat}
  {}{}
}%
\makeatother
\begin{document}
\title{Enhanced heat transfer in a two-dimensional serpentine micro-channel using elastic polymers} 

\author{Himani Garg}
\author{Lei Wang}%
\email{himani.garg@energy.lth.se}
 \affiliation{$^1$Lund University$,$ Department of Energy Sciences $,$ P$.$O$.$ Box 118$,$ SE-22100 Lund$,$ Sweden}
\begin{abstract}
\section*{Abstract}
{
Elastic turbulence has emerged as a promising method to enhance heat transfer performance at the microscale. However, previous studies have mainly focused on the overall convective heat transfer performance in curved channels, overlooking the fact that the chaotic flow intensity varies along the streamline, leading to diverse local heat transfer characteristics. In this study, we systematically investigate the hydraulic and thermal properties of a dilute polymer solution under elastic turbulence conditions, where the inflow conditions exhibit a vanishing Reynolds number ($Re$) and high Weissenberg number ($Wi$), enabling a comprehensive understanding of the influence of polymers on the system. Through extensive direct numerical simulations using Rheotool, a two-dimensional curvilinear channel flow of an Oldroyd-B viscoelastic fluid is analyzed. By examining the variations of the friction factor and Nusselt number along the serpentine channel, we unveil both global and local characteristics of elastic turbulence. Based on the $Wi$ value, we identify three distinct regimes. In the first regime ($0 < Wi \leq 3$), we observe approximately 10\% heat transfer enhancement accompanied by a roughly 5\% reduction in the friction factor compared to laminar flow, known as polymer-induced thermal conductivity enhancement. In the second regime ($3 < Wi \leq 5$), we observe a steep linear increase in heat transfer (around 30\%) at the expense of up to 15\% enhanced friction factor. Finally, in the fully developed elastic turbulence regime ($Wi > 5$), we observe a remarkable heat transfer enhancement of up to 60\% along with a reduced friction factor. The significant enhancement of heat transfer with increasing $Wi$ can be attributed to the intensifying elastic instability resulting from the balance between normal stresses and streamlined curvatures.}
\end{abstract}

\maketitle

\section{Introduction}
\label{Sec: 1}
{In recent decades, microfluidics, with its ability to manipulate fluids at the microscale, has emerged as a promising technology with diverse applications in the field of energy. This technology offers numerous benefits, including compactness, high surface-to-volume ratio, precise control over fluid behavior, and the potential for efficient heat transfer. Microfluidic systems play a crucial role in various energy-related applications such as microreactors for catalysis~\cite{RENKEN201047}, microscale heat exchangers~\cite{FAGHRI2006421}, cooling of electronics~\cite{ZHANG2021100009}, fuel cells~\cite{BARGAL2020635}, and microscale heat pipes~\cite{FAGHRI2006421}. These systems provide improved performance, enhanced energy efficiency, and effective thermal management. 
However, with the reduction in device size, the challenges associated with thermal control and efficient heat removal have become increasingly critical. This is primarily due to the limitations imposed by the laminar flow regime, i.e., low Reynolds number ($Re$) flows dominating over inertial forces, which hinders convective heat transfer. 
While laminar flows in microfluidic systems provide precise control over flow velocity and pattern~\cite{li2015experimental}, they suffer from limitations in mass transfer perpendicular to the main flow direction and poor heat transfer in microscale cooling systems~\cite{Burghelea2004}. To overcome these challenges, various techniques have been developed to enhance heat transfer with Newtonian fluids. These techniques include the use of vortex promoters~\cite{meis2010heat}, periodic expansion-constriction structures~\cite{bryce2010abatement}, rough wall surfaces~\cite{shen2006flow}, and changing working mediums~\cite{pimenta2013heat,kiyasatfar2016laminar}. The primary objective of these methods is to disrupt the low $Re$ flows and increase the contact interfaces between fluids of different concentrations or temperatures. By introducing flow disturbances and promoting mixing, these techniques aim to improve convective heat transfer in microfluidic systems and enhance the overall heat transfer efficiency. }

Remarkably, despite the limitation of vanishing $Re$ in microfluidics, recent studies have demonstrated that viscoelastic fluids, particularly shear-thinning fluids, can induce turbulence-like states at the microscale due to the coiling and stretching of dissolved polymers, which is referred to as Elastic Turbulence \cite{Groisman2000,Groisman2004,Steinberg2021}. {When long-chain polymer or surfactant molecules traverse curved streamlines, the viscous shear forces cause the stretching of these molecules. This stretching generates a non-linear hoop stress that acts as a radial disturbance within the flow field. The intriguing interaction between the elastic stress and strains amplifies these disturbances, eventually resulting in chaotic motion. The coupling between elastic stress and strain dynamics plays a crucial role in the generation and amplification of perturbations within the flow, leading to complex and unpredictable fluid behavior.} This phenomenon occurs in viscoelastic flows with an extremely low $Re$ and high Weissenberg number ($Wi$), which is the ratio of elastic to viscous forces, as initially reported by Groisman and Steinberg in 2000 \cite{Groisman2000}. Since their discovery, elastic instabilities have been experimentally induced and characterized in various flows within standard geometries, including Taylor-Couette flows \cite{Larson1990}, von Kármán swirling flows \cite{Burghelea2004, Groisman2004}, other curved flows in serpentine channels \cite{Abed2016}, {straight channel with cylindrical obstacles~\cite{li2010creation, Arratia2017}, or in designed porous media~\cite{browne2021elastic}. In this regime, the viscoelastic flows display characteristics similar to turbulent flows, such as power-law decay of the velocity power spectra with an exponent greater than three, highly non-Gaussian distributions of velocity gradients indicating intermittency, and dramatic increase in the flow resistance usually contributing to mass and heat transfer enhancement at the macroscale. These findings suggest that elastic turbulence has promising applications in microfluidics as it enhances phenomena such as mixing and heat transfer.}

More recently, it has been found that heat and mass transfer enhancement can be achieved in microscale flows with simple 2D geometries, including parallel disks and serpentine channels, due to the onset of chaotic-like flows \cite{Groisman2000,Burghelea2004,Burghelea2007,Whalley2015}. Traore et al. \cite{Traore2015} first reported up to a four-fold enhancement of heat transfer efficiency in the regime of elastic turbulence in von Kármán swirling flows, comparable to the efficiency in inertial turbulence at $Re=1,600$. This finding was later confirmed by Ligrani et al. \cite{Ligrani2018}. Moreover, experiments carried out by Whalley et al. \cite{Whalley2015} and Abed et al. \cite{Abed2016} in microscale serpentine channels with a constant temperature boundary condition demonstrated an increase in heat transfer coefficients by 300\%. These studies reveal that various geometries can benefit from the onset of elastic instability and elastic turbulence at extremely low $Re$ to enhance heat transfer performance. \textcolor{black}{However, accessible geometries are predominantly limited to planar flows, for instance, swirling flows between two rotating parallel disks or cylinders, flow in two parallel plates with an array of cylinders, contractions, and cross-slots, etc.}. Furthermore, curved channels can be considered simple 2D planar geometries where the flow evolves in two dimensions, and the pressure drop is relatively high due to sharp turning points. This significant pressure drop is considered one of the three main features of {elastic turbulence} that indicate the complete transition from elastic instability to {elastic turbulence}, as systematically summarized by Steinberg in a recent review \cite{Steinberg2021}. {Another recent multi-author review article~\cite{datta2022perspectives}, based on the virtual workshop on viscoelastic flow instabilities and elastic turbulence organized by the Princeton Center for Theoretical Sciences, also provides a state-of-the-art summary of the various challenges in this broad area. While previous experiments have focused on passive scalar transport facilitated by elastic instabilities and turbulence, the evaluations have primarily been limited to statistical characteristics. Consequently, there remains a lack of detailed information regarding how the flow dynamics of elastic turbulence impacts the heat transfer process. The absence of corresponding information hinders our understanding of crucial aspects, such as the mechanisms behind mass and heat transfer enhancements, the scaling of the Nusselt number ($Nu$) and $Wi$, the interaction between heat transport and the velocity field of elastic turbulence, and more. Obtaining such information experimentally poses challenges, including acquiring insights into the features of stretched polymers and a deeper understanding of heat convection driven by the disordered fluid motion. These limitations have significantly constrained our current understanding of these phenomena.}

{To compensate for experimental limitations, numerical simulations offer valuable reference information. For instance, Berti et al. \cite{Berti2008} and Garg et al. \cite{Garg2018,Garg2021} conducted direct numerical simulations using the Oldroyd-B constitutive model to investigate elastic turbulence in two-dimensional Kolmogorov flow. \textcolor{black}{The studies by Garg \textit{et al.}~\citep{Garg2021}} have successfully reproduced elastic turbulence in a simplified setup and revealed that Taylor's frozen-field hypothesis does not fully apply, despite the presence of a well-defined mean flow. Grilli et al. \cite{grilli2013transition} employed the Oldroyd-B constitutive model to simulate viscoelastic fluid flow through a cylinder array in a channel, demonstrating that curved streamlines were not a prerequisite for the development of elastic turbulence. Poole et al. \cite{poole2013viscoelastic} conducted numerical simulations of viscoelastic secondary flow in a serpentine channel using the Oldroyd-B model, exploring a range of $Wi$ from $0$ to $0.6$. \textcolor{black}{It should be noted that despite the efforts put into simulations, the relevance of scaling relations dictating the interaction between heat transport and the velocity field of elastic turbulence, as obtained from 2D simulations, is limited when applied to the 3D flows studied experimentally. This limitation arises because the base flow structure before the onset of elastic instability differs between the two cases.} Although there have been a few numerical studies analyzing elastic turbulence regimes and heat transfer enhancement, comprehensive investigations remain scarce \cite{Buel2022,Canossi2020,Song2022}. Moreover, to the best of our knowledge, only a few numerical works have provided in-depth analyses of the elastic turbulence regime \cite{Garg2021} and its impact on heat transfer enhancement \cite{Li2017}.
Li et al. \cite{Li2017} focus on the heat transfer characteristics in 3D serpentine channels of "Omega" shape that are different from the ones considered in the present study. Their focus is limited to the global characteristics. 
However, these computational results on elastic turbulence applications are predominantly focused on relatively low $Wi$, and the numerical exploration of heat transfer accompanied by elastic turbulence remains incomplete. The primary challenge in simulating realistic geometries lies in the high $Wi$ problem, which leads to breakdowns in numerical strategies when solving various constitutive equations for viscoelastic fluids at relatively high $Wi$, often associated with Hadamard instabilities \cite{sureshkumar1995effect}. It is generally recognized that Hadamard instabilities stem from the loss of the symmetric positive definiteness property of the conformation tensor and also from the deviation between polynomial fitting and exponential profiles of the conformation tensor at high deformation rates. Many efforts have been paid to deal with these instabilities, such as the decomposition algorithm for conformation tensor~\cite{vaithianathan2003numerical} or the velocity gradient decomposition of the conformation tensor, also known as the log-conformation reformulation method~\cite{Fattal2004}.}

{To the best of our knowledge, there is no report of a numerical study focusing on local heat transfer enhancement induced by elastic turbulence in the literature. In the present work, for the first time, we realize the numerical simulations of elastic turbulence and its subsequent impact on heat transfer in a two-dimensional curved channel, inspired by experiments~\cite{zilz2012geometric}. These simulations were performed using the Oldroyd-B model implemented in the open-source software OpenFOAM. \textcolor{black}{While it is known that linear elasticity models, such as Oldroyd-B, typically
underestimate experimentally measured elastic stresses~\cite{liu2010molecular} and do not account for shear-thinning~\cite{Whalley2015,Traore2015}, it has been argued that the main features of elastic turbulence are relatively independent of the specific rheological model details~\cite{plan2017lyapunov,Garg2018,Garg2021,fouxon2003spectra}. Nevertheless, the effect of rheological model choice on heat transfer properties should not be overlooked. The shear-thinning nature of the fluid influences the base flow topology (particularly at $W i \leq Wi_c$ ) and likely the primary flow characteristics observed in the regime of elastic turbulence. Rheological models that consider shear-dependent viscosity effects, such as FENE models, may introduce additional dynamic couplings between the flow and the heat transfer. Additionally, while the range of temperatures studied experimentally shows weak variations in heat capacity and thermal diffusivity, the rheological properties, particularly shear viscosity and the characteristic relaxation time, vary significantly with temperature. The thermal dependence of fluid properties has been neglected in this study, which may be a consideration for future simulations. It is worth noting that all experiments were conducted in a three-dimensional setting. Comparing heat transfer perspectives in curvilinear channels between 3D and 2D cases is challenging because the base flow structure significantly differs before the onset of elastic instability in the two cases. Therefore, for various technical reasons, a direct comparison with experimental data is not attempted here. Instead, our primary focus is on establishing the foundations for advancing our understanding of 3D setups.} The paper is organized into five main sections, with the remaining sections following as outlined: Sec. \ref{Sec: 2} and Sec. \ref{Sec: 3} describe the methodology used for the numerical simulations and provide detials regarding boundary conditions and dimensionless numbers used in the analysis. Sec. \ref{Sec: 4} presents the results and discussions, and finally, the conclusions are drawn in Sec. \ref{Sec: 5}.}
\section{Numerical Models and Methods}
\label{Sec: 2}
 \begin{figure}[!ht]
\begin{center}
\includegraphics[width=1\linewidth]{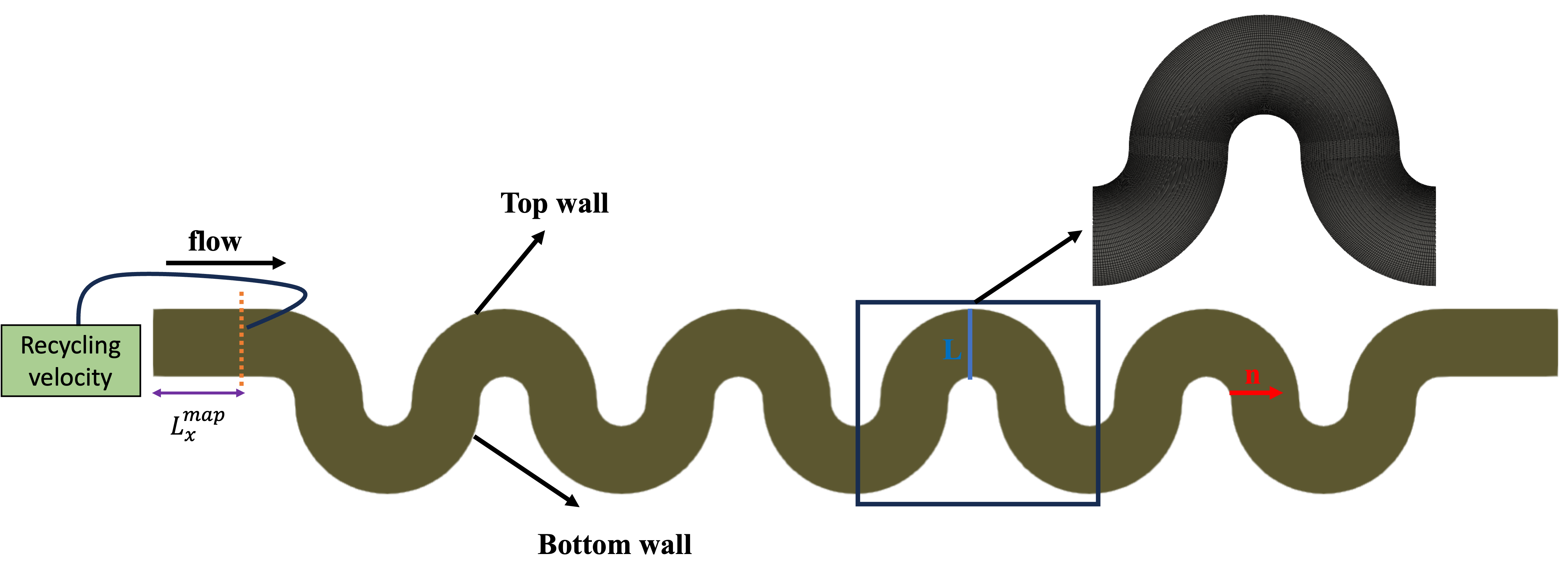}
\caption{{Schematic of the microfluidic domain with constant inner and outer curvatures. $\bm{x}$ is the primary flow direction and $\bm{n}$ is the wall-normal direction where the origin is taken as the inner edge of each loop. The location of the line probe used for mean flow profiles is also indicated, $L = x/W=6.9$.}}
\label{fig: -1}
\end{center}
\end{figure}
In this study, we investigate a 2D viscoelastic fluid flow within a serpentine micro-channel, as depicted in {Duclou{\'e} \textit{et al.}} \cite{ducloue2019secondary}, shown in Fig.~\ref{fig: -1}. The fluid is assumed to be incompressible and inertialess.\textcolor{black}{The micro-channel consists of three parts: first planar channel of length  $L_x^{in}=190 \mu$m followed by $N=9$ connected half-circular rings, with inner and outer radii of {$R_1=40\mu$m and $R_2=150\mu$m, respectively and finally another planar channel of length  $L_x^{out}=190\mu$m}. The channel height is $W=R_2-R_1=110\mu$m, and the geometric aspect ratio $a=R_1/W=0.36$} is relevant to the primary elastic instability \cite{Zilz2012.}. The micro-channel has inlet and outlet boundaries along the $ x-$ direction and top and bottom walls along the $y-$direction. Due to the small size of the simulated channel, the flow is dominated by nonlinear elasticity, with negligible inertial effects due to very low $Re$. \textcolor{black}{ In experiments, the channel is very long (roughly 200 turns) for the flow to fully develop, particularly in the regime of high $Wi$ numbers. However, when simulations are performed, due to small time step limitation, long geometries are prohibitive. We have estimated the development length for the laminar base flow using the following correlation $L_{dev}/W=0.06Re$, which implies $L_{dev}=44$nm. In our case, $L_{dev} \ll  L_{x}$, where $L_x=2.2$mm is the total channel length. So it is safe to say that the considered length is sufficient for the flow to be fully developed at least in the regime of small $Wi$. For the laminar flow case, we have also plotted the profiles at different locations in the channel (not shown here), and as expected these profiles converge to a single master curve as $L_{dev}$ is significantly small compared to $L_x$. Furthermore, to obtain numerical results independent of the entrance effects, we generated fully developed flow in the regime of elastic turbulence by recycling the velocity filed at the inlet from an arbitrary plane located at a distance $L_x^{map} =150\mu$m downstream of the inlet. The cyclic boundary conditions are imposed for the inlet and the arbitrary plane downstream. The recycled velocity then enters the rest of the channel continuously and flows toward the outlet using mapping boundary conditions in OpenFOAM. Note that recycling is only done for large values of $Wi$.} We describe the evolution of the velocity field $\bm{u} (\bm{x}, t)$ at position $\bm{x}$ and time $t$ using the mass, momentum, and energy conservation equations, assuming the continuum hypothesis and incompressible flow conditions:
\begin{equation}
 \nabla \cdot \bm{u}=0,
    \label{Eq: 1}
\end{equation}
\begin{equation}
\rho \left[ \frac{\partial \bm{u}}{\partial t} + \left( \bm{u} \cdot \nabla\right) \bm{u}\right] =  -\nabla p + \nabla \cdot \bm{\tau},
    \label{Eq: 2}
\end{equation}
\begin{equation}
\frac{\partial T}{\partial t} + \bm{u}\cdot \nabla T = \alpha \nabla^2T,
    \label{Eq: 3}
\end{equation}
where  $\rho$ is the fluid density, $p$, the fluid pressure, $\bm{\tau} = \bm{\tau}_s+\bm{\tau}_p$ is the fluid stress tensor which consists of both solvent ($\bm{\tau}_s$) and elastic ($\bm{\tau}_p$) solute contribution, $T$, the fluid temperature, and $\alpha$, the fluid thermal diffusivity assuming constant $c_v$. 

In the framework of the Oldroyd-B model \cite{Bird1987}, $\bm{\tau}_s = 2\eta_s\textbf{S}$, where $\eta_s$ is the zero-shear dynamic viscosity of the solvent, $\textbf{S}=\frac{1}{2}\left(\nabla \bm{u} + (\nabla \bm{u})^T\right)$ is the strain-rate tensor, and $\bm{\tau}_p$ is expressed by the following constitutive equation following Oldroyd, \cite{Oldroyd1950} 
\begin{equation}
\bm{\tau}_p+\lambda \left[ \frac{\partial \bm{\tau}_p} {\partial t} + \nabla \cdot (\bm{u\tau}_p) - \left(\nabla \bm{u} \right)^T \cdot \bm{\tau}_p - \bm{\tau}_p \cdot \left(\nabla \bm{u} \right)\right] = 2\eta_p\textbf{S},
    \label{Eq: 4}
\end{equation}
where $\lambda$ is the largest polymer relaxation time and $\eta_p$ is the polymer contribution to viscosity. An important parameter is the viscosity ratio, $\beta = \eta_s/\left(\eta_s+\eta_p \right)$, inversely proportional to the polymer concentration. 
{Equation \ref{Eq: 4} can also be written in terms of conformation tensor, $\bm{C} = \frac{\lambda}{\eta_p} \tau_p + \textbf{I}$, as follow:
\begin{equation}
 \frac{\partial \bm{C}} {\partial t} + \bm{u} \cdot \nabla \bm{C} - \left(\nabla \bm{u} \right)^T \cdot \bm{C} - \bm{C} \cdot \left(\nabla \bm{u} \right) = \frac{1}{\lambda}(\textbf{I}-\bm{C}),
    \label{Eq: conformation tensor -5}
\end{equation}
where $\textbf{I}$ is the identity matrix.}
For a given value of $\beta$, the main control parameters of the dynamics specified by Eq. (\ref{Eq: 2}) and Eq. (\ref{Eq: 4}) are the Reynolds number, $Re = \rho U_{{max}} W /\left(\eta_s+\eta_p\right)$, and Weissenberg number, $Wi=\lambda U_{{max}}/W$, where $U_{{max}}$ is the maximum velocity intensity at the inlet.

\textcolor{black}{For the sake of generality, the governing equations Eqs.~\ref{Eq: 1} to ~\ref{Eq: 3} and Eq.~\ref{Eq: conformation tensor -5} are converted to the dimensionless form:}
\textcolor{black}{\begin{equation}
 \nabla^+ \cdot \bm{u}^+=0,
    \label{Eq: 6}
\end{equation}}
\textcolor{black}{\begin{equation}
\frac{\partial \bm{u}^+}{\partial t^+} + \left( \bm{u}^+ \cdot \nabla^+ \right) \bm{u}^+ = -\nabla^+ p^+\frac{\beta}{Re}\nabla^{+2}\bm{u}^++\frac{(1-\beta)}{Re\ Wi} \nabla^+ \cdot \bm{C},
    \label{Eq: 7}
\end{equation}}
\textcolor{black}{\begin{equation}
\frac{\partial T^+}{\partial t^+} + \bm{u}^+\cdot \nabla^+ T^+ = \frac{1}{Re\ Pr} \nabla^{+2}T^+,
    \label{Eq: 8}
\end{equation}}
\textcolor{black}{\begin{equation}
 \frac{\partial \bm{C}} {\partial t^+} + \bm{u}^+ \cdot \nabla^+ \bm{C} - \left(\nabla^+ \bm{u}^+ \right)^T \cdot \bm{C} - \bm{C} \cdot \left(\nabla^+ \bm{u}^+ \right) = \frac{1}{Wi}(\textbf{I}-\bm{C}),
    \label{Eq: 9}
\end{equation}}
\textcolor{black}{where, $\nabla^+ = W \ \nabla$, $\bm{u}^+ = (u_x,u_y)/U_{max}$, $t^+=t\ U_{max}/W$, $T^+ = (T-T_0)/T_0$, and $p^+=p/\left( \rho U^2_{max} \right)$}.
\section{Discrete schemes and boundary conditions} 
\label{Sec: 3}
Equations (\ref{Eq: 1}) and (\ref{Eq: 3}) were solved using the open-source numerical solver RHEOTOOL \cite{rheoTool}, which was developed in the framework of OpenFOAM \cite{Weller1998}. This solver is based on the finite-volume discretization and utilizes the log-conformation technique \cite{Fattal2004} to control numerical instabilities associated with high $Wi$ values. Notably, polymer-stress diffusion was not considered. The convective fluxes were discretized using the third-order accurate CUBISTA scheme for momentum and energy conservation equations \cite{Alves2003}, while the convective term in the conformation tensor transport equation was discretized using a second-order accurate scheme. 

{As for the boundary conditions, a parabolic velocity profile and uniform temperature distribution are set at the inlet as
\begin{equation}
u(y) = U_{{max}}(1-y^2),  \hspace{0.3em} T =T_{0}, \hspace{0.3em} \textnormal{and} \hspace{0.3em} \frac{\partial p}{\partial x}=0,
    \label{Eq: inlet BC}
\end{equation}
was imposed at the inlet, with zero-Neumann boundary conditions specified for the pressure field. The natural flow out boundary condition is given at the outlet with the developed boundary condition for the temperature field as:
\begin{equation}
p = 0, \hspace{0.3em} \frac{\partial T}{\partial x}=0, 
    \label{Eq: outlet BC}
\end{equation}
The velocity field at the walls was subjected to a no-slip condition, while a linear extrapolation technique was employed for the conformation tensor. A Dirichlet condition was utilized for the temperature, $T_w$, which was set to be higher than the inlet temperature. The initial values for the conformation tensor and elastic stress fields were set as a unit tensor and zero tensor, respectively, representing the coiled state.}

For the Oldroyd-B type of viscoelastic fluids and channel geometry, the heat transfer performance can be determined by four dimensionless parameters, i.e., $Nu \propto f(Wi, Re, Pr, \beta)$. In this study, the values of \textcolor{black}{$\beta=0.8$, $Re$, and $Pr=1000$ were held constant to solely investigate the effect of varying $Wi$ on flow and heat transfer properties. Thus, the Nusselt number only depended on $Wi$. 
Despite the vanishing $Re$ number at {O($10^{-3}$)}, the Peclet number, $Pe=6.7$, which is the product of $Re$ and $Pr$ number, is set high, i.e., convection and diffusion are of the same order of magnitude in heat transfer.} 
$Wi$ varied from 0 to 20, reflecting the transition from viscosity-dominated to elasticity-dominated flow. Here, $Wi=0$ represents Newtonian laminar flow. The time step for the simulations was chosen as $\Delta t = \min{\left( \lambda, W/U, W^2/\nu \right)}$, {where $U=2U_{max}/3$ is the mean inlet velocity},  $W/U$ is the convective time step, and $W^2/\nu$ is the diffusive time step. \textcolor{black}{We estimated the effect of viscous dissipation in the micro-serpentine channel. Theoretically, viscous dissipation is represented by $\phi = \mu \left( \partial u_i/\partial x_j + \partial u_j/\partial x_i\right) \partial u_i/\partial x_j$, in our case, we use the rough estimation in which $\phi=\mu \left(U/W\right)^2$. Meanwhile, we calculated heat transfer into the channel per unit volume, that is, $\rho c_p U \left( T_{out}-T_{in}\right)/L_x$, where $T_{out}$ is the outlet temperature. We estimated the ratio of viscous dissipation to the heat transfer rate per unit volume, and we found that it is extremely low because it is less than $5 \times 10^{-8}$. Therefore, the viscous dissipation is negligible in our case.}
\section{Results and Discussions}
\label{Sec: 4}
 \begin{figure}[!ht]
\begin{center}
\begin{subfigure}[b]{0.45\textwidth}
\includegraphics[width=\textwidth]{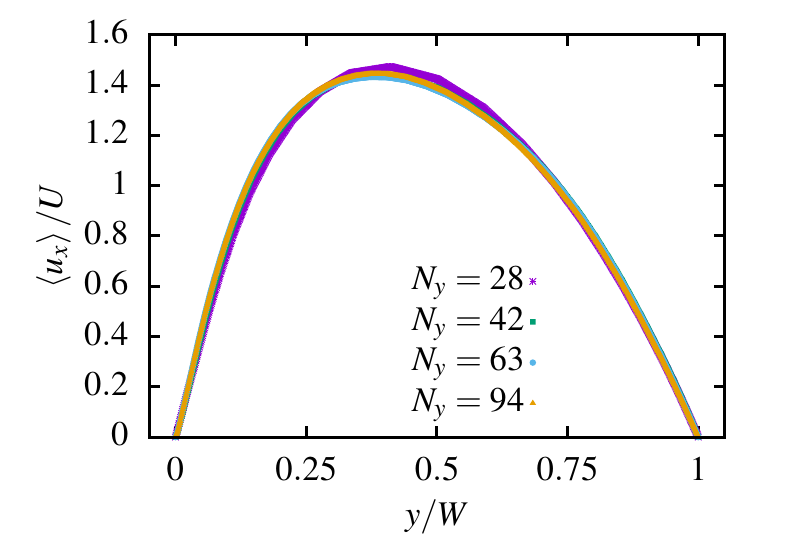}
\caption{}
\label{fig: 0a}
\end{subfigure}
\begin{subfigure}[b]{0.45\textwidth}
\includegraphics[width=\textwidth]{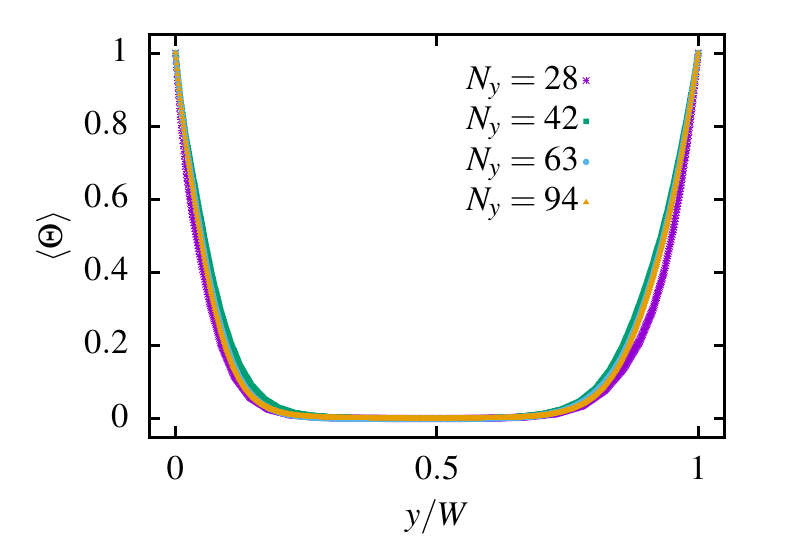}
\caption{}
\label{fig: 0b}
\end{subfigure}
\caption{{Normalized (a) mean flow profile, $\left< u_x/U \right>$ and (b) mean temperature profile, $\left<\Theta\right>$ at $L = x/W=6.9$ for five sets of mesh resolution. Here $Wi=5$ is fixed.}}
\label{fig: 0}
\end{center}
\end{figure}
{\subsection{Mesh independence}}
In the present study, a comprehensive analysis was conducted to assess the mesh independence of the targeted curvilinear micro-serpentine channel, with a specific focus on the intermediate $Wi=5$ case. \textcolor{black}{The computational domain was discretized using $N_x = 4480$ uniform intervals along the longitudinal direction. To ensure accurate representation near the wall in the cross-section, four sets of grid distributions, namely $N_y \times N_z$ = $28 \times 1$, $42 \times 1$, $63 \times 1$, and $94 \times 1$, were employed. The mesh refinement technique was implemented to enhance the resolution in these regions, as shown in Fig.~\ref{fig: -1}. The total number of cells in these meshes varried from 125,440 to 421,110.}
To maintain computational stability, all simulation cases were performed using a dimensional time step of $\Delta t$. The resulting dimensionless mean velocity profile, denoted as $\left< u_x/U\right>$, and temperature profile, represented by $\left< \Theta\right> = (T-T_0)/(T_w-T_0)$, at position $L = x/W = 6.91$ (see Fig. \ref{fig: -1}) are illustrated in Fig. \ref{fig: 0}.
The analysis reveals a clear trend of increasing accuracy and convergence as the mesh is refined. Considering both the accuracy of the different mesh configurations and the computational efficiency, a cross mesh with dimensions of $42 \times 1$ was chosen as the optimal choice for this study.

\begin{figure*}[!ht]
\begin{center}
\begin{subfigure}[b]{0.45\textwidth}
\includegraphics[width=\textwidth]{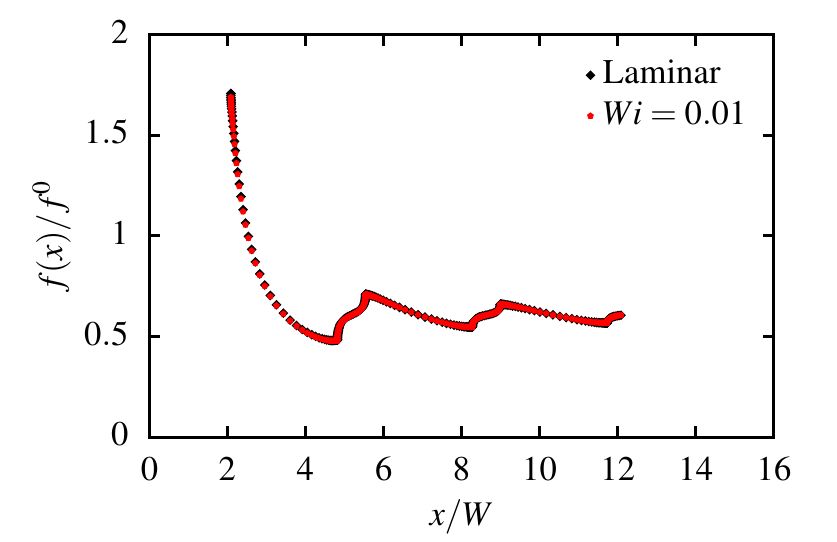}
\caption{}
\label{fig: 1a}
\end{subfigure}
\begin{subfigure}[b]{0.45\textwidth}
\includegraphics[width=\textwidth]{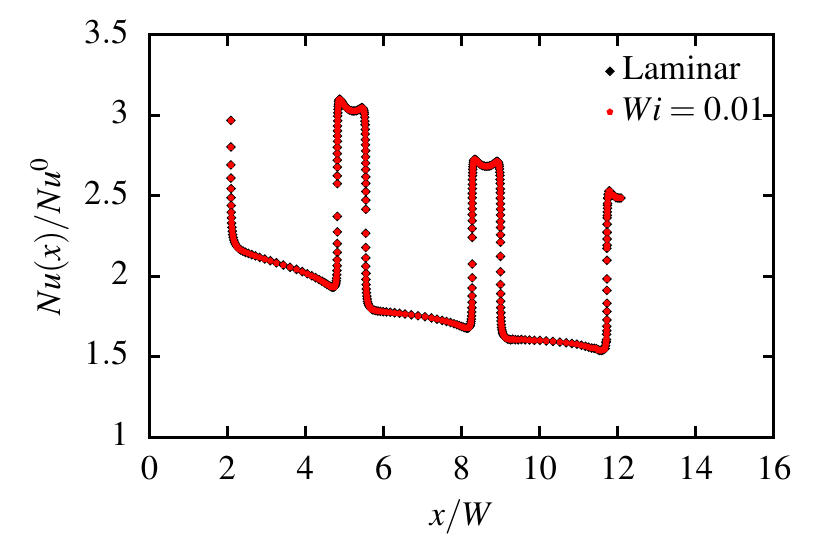}
\caption{}
\label{fig: 1b}
\end{subfigure}
\caption{Effect of polymers on the local friction factor and heat transfer: dependence of the local friction factor, $f(x)$, and $Nu$ on $x$ along the full channel.}
\label{fig: 1}
\end{center}
\end{figure*}

\textcolor{black}{To further ensure that the chosen mesh size is accurate enough such that results are not biased by numerical diffusion we have estimated the characteristic length associated to the temperature diffusivity, $L_c=\sqrt{\alpha t_D}$, where $\alpha$ is the thermal diffusivity and $t_D$ is the diffusion time step. In present simulations, the value of $L_c=3.5\mu$m and the mesh size close to the wall is $0.68\mu$m. This estimate shows that the chosen mesh size is significantly fine enough to ensure well resolved thermal problem.
Lastly, we calculate the surface average value of friction factor, $f$ and Nusselt number, $Nu$, as a function of mesh resolution, see Table~\ref{table 1}. The relative error for $f$ and $Nu$ with respect to the finest mesh are also shown in Table~\ref{table 1}. As expected with mesh refinement the relative error for both $f$ and $Nu$ number reduces and converges towards finer mesh. Maximum of error (14\% for $f$ and 34\% for $Nu$ number) is obtained for the G1 grid. However, for all other grids G2-G4, the results are close and the maximum relative error is between 1\% to 3\%. As a trade-off between computational time and accuracy, we decided to continue with grid G2, and perform further parametric study.}
\begin{table}[]
\begin{tabular}{|l|l|l|l|l|l|l|}
\hline
Grids &Ncells & $N_y$ & $f$  & $Nu$    & $f_{error}\%$ & $Nu_{error}\%$  \\ \hline
G1 & 125440 & 28  & 6142 & 18.49 & 13.95  & 34.35\\ \hline
G2 & 188160 & 42  & 7119 & 13.34 &1.01  &3.12 \\ \hline
G3 & 282240 & 63  & 7210 & 13.40 & 0.26 & 2.70\\ \hline
G4 & 421120 & 94  & 7137 & 13.77 & 0.00 & 0.00\\ \hline
\end{tabular}
\caption{Estimate of surface averaged values of friction factor, $f$, Nusselt number, $Nu$ and their relative errors,$f_{error}\%$ and $Nu_{error}\%$, respectively.}
\label{table 1}
\end{table}
{\subsection{Laminar flow validation}}
Before performing any parametric study, we investigated the benchmark laminar flow case mainly by looking at the local measurements of steady-state pressure drag and heat transfer along the curvilinear channel, shown in Fig. \ref{fig: 1}. The local friction factor, $f(x)$, is defined as
\begin{equation}
  {  f(x) = \frac{2\Delta p}{ L_x }\frac{W}{\rho U^2},}
\end{equation}
where $\Delta p/L_x$ is the local pressure gradient. Here, $\Delta p$ is the pressure drop with respect to the inlet pressure, while $L_x$ is the local distance. 
While the local estimation of the Nusselt number, $Nu(x)$, given by
\begin{equation}
    Nu(x)=\frac{h(x) W}{k}
\end{equation}
 where $h(x)=-k \left( \frac{dT(x)}{dn}\right)_w/\left(T_w-T_b(x) \right)$ and $h=\dot{m}c_p\left(T_2-T_1 \right)/\left(A_s \Delta T_{{avg}}\right)$ are the local and global estimates of the convective heat transfer coefficient, respectively, $W$, the height of the micro-channel, $k$, the thermal conductivity of the fluid, $\left( \frac{dT(x)}{dn}\right)_w$, the local wall normal temperature gradient, $T_w$, the wall temperature, $T_b(x)$, the local bulk temperature, $\dot{m}$, the mass flow rate, $c_p$, the specific heat capacity at constant pressure, $T_1$ and $T_2$, the inlet and outlet temperatures, $A_s$, the area of the heated surface, and $\Delta T_{{avg}}$, the average temperature difference.  
   \begin{figure*}[!ht]
\begin{center}
\includegraphics[width=1\linewidth]{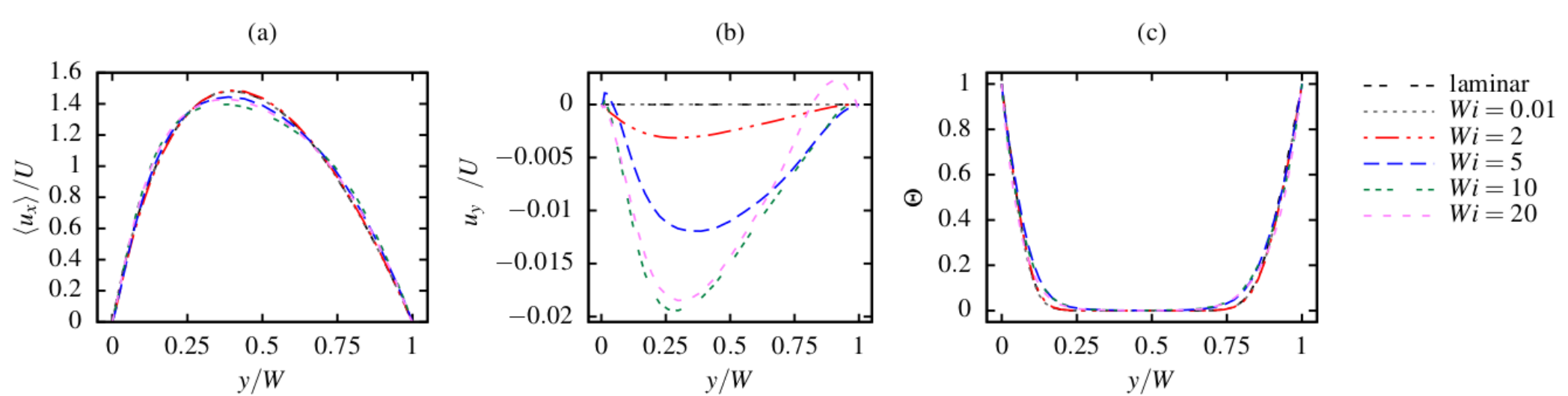}
\caption{Effect of $Wi$ on the normalized local mean flow profile (a) streamwise velocity component, (b) wall-normal velocity component,  and (c) mean temperature profile at a selected position, $x/W=6.9$, along the wall-normal direction.}
\label{fig: 2new}
\end{center}
\end{figure*}

\textcolor{black}{In addition to the benchmark case, we present results for $Wi=0.01$ (see Fig. \ref{fig: 1}). The results are normalized by the average value of friction factor, $f^0=0.64/Re$, and Nusselt number, $Nu^0=4.36$, calculated for the benchmark laminar Newtonian plane channel case. For such a low value of $Wi$, the flow is considered laminar since all flow fluctuations are negligible (as demonstrated later). We assume that the values of $f(x)$ and $Nu(x)$ are similar to those of the benchmark case. In laminar flow with the same $Re$ number and $Wi=0.01$, the friction factor must be the same because $f$ is inversely proportional to $Re$ (as shown in Fig. \ref{fig: 1a}). The local $Nu$ number results for $Wi=0.01$ also converge to the laminar Newtonian flow solution as expected (see Fig. \ref{fig: 1b}). This further validates that the chosen mesh and numerics are reliable.}

\textcolor{black}{Additionally, the results shown in Fig.~\ref{fig: 2new} indicate periodic variations throughout the channel surface. This periodicity in the results is closely linked to the serpentine channel geometry, which consists of repeatable loops, i.e., a pair of upper and bottom half-circular rings. As the fluid flows through the channel, it encounters these rings repeatedly and each ring introduces a change in the flow velocity, pressure, and heat transfer properties due to the centrifugal forces. More details on the relationship between the local behavior of $Nu(x)$ number and serpentine-channel geometry are provided later in Sec.~\ref{Sec: local flow properties}. Apart from the periodicity, we also observed that the absolute values of the $f(x)$ and $Nu(x)$ numbers decrease along the channel surface as we move downstream. For $Nu(x)$ number, this is more visible and is related to diffusion as thermal boundary layer thickens along the channel. Further simulations with an even longer channel (probably 50 turns) are warranted to explore how the $Nu(x)$ number changes over the length. On the other hand, for $f(x)$ number, the magnitude decrease with channel length is also present but less dramatic and seems to converge towards a plateau faster than $Nu(x)$ number. One possible reason for that is the different “adaptation” response time of the kinematic and thermal boundary layers to the periodic perturbations induced by the geometry.}

\subsection{Mean flow profiles}
Figure \ref{fig: 2new} presents the time-averaged streamwise velocity component, instantaneous wall-normal velocity component, and time-averaged temperature profile at a location of $L = x/W=6.9$. In this plot, $y$ represents the wall-normal direction, and $x$ denotes the primary velocity direction. The figure is plotted in a way that the zero $y$ position corresponds to the inner edge at $L = x/W=6.9$. In Fig. \ref{fig: 2new}(a), the streamwise velocity component is shown. Firstly, it is noticeable that the Newtonian simulations and the small Weissenberg regime (black, grey, and red dashed lines) agree well, but a slight asymmetry towards the inner wall is present. Similar observations have been made and discussed in previous studies \cite{zilz2012geometric}. Secondly, the effect of elasticity on the main velocity component is relatively stable, but upon examination of the profiles around the maximum of $\left<u_x\right>$ (not shown in this figure), it is apparent that elasticity reduces this asymmetry by shifting the peak velocity towards the center of the channel. Figure \ref{fig: 2new}(b) shows the wall-normal velocity component. In the case of Newtonian fluids, this component is negligible, consistent with the theoretical predictions for a duct with a constant cross-sectional area and curvature, and without any inertia, as stated in \cite{lauga2004three}. However, when a polymer solution is introduced, an increasing magnitude of wall-normal velocities from the inner wall toward the outer wall can be observed as $Wi$ increases. Note that, like the streamwise velocity component, these transverse profiles are also asymmetric, with a peak closer to the inner wall. This is attributed to a larger shear rate near the inner wall, which drives the secondary flow through the streamwise normal stresses and streamline curvature. As the driving force increases with the shear rate, the higher shear rate at the inner wall leads to an asymmetry in the distribution of the secondary flow. We then turn to the mean normalized temperature profile, $\Theta$, shown in Fig. \ref{fig: 2new}(c).
As anticipated, there is a good agreement between Newtonian simulations and the small $Wi \leq 2$ regime. At the location $L= x/W = 6.9$, the thermal boundary layer is still developing for all considered Wi values, as the temperature in the central region remains equal to the inlet temperature. Unlike the velocity profiles, the temperature profiles are almost symmetrical regardless of the $Wi$ values. \textcolor{black}{This is related to the $Pe=6.7$ number which implies that heat convection has a comparable effect to molecular heat diffusion.} When $Wi > 5$, the fluid temperature close to the wall is higher than for $Wi = 0.01$ and $2$, indicating greater heat transfer has been achieved.

 \begin{figure}[!ht]
\begin{center}
\begin{subfigure}[b]{0.45\textwidth}
\includegraphics[width=\textwidth]{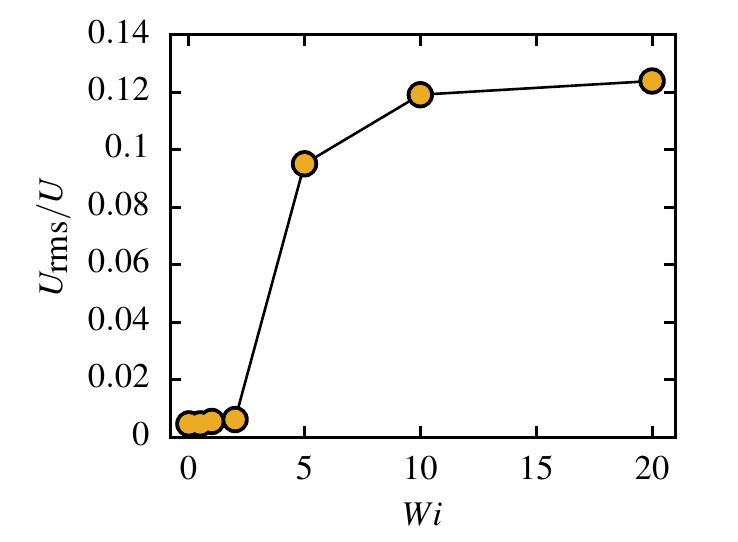}
\caption{}
\label{fig. 2a}
\end{subfigure}

\begin{subfigure}[b]{0.45\textwidth}
\includegraphics[width=\textwidth]{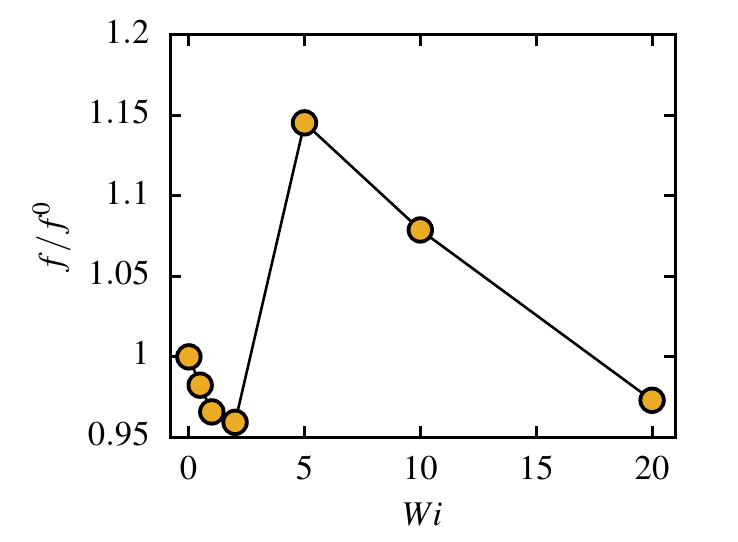}
\caption{}
\label{fig. 2b}
\end{subfigure}

\begin{subfigure}[b]{0.45\textwidth}
\includegraphics[width=\textwidth]{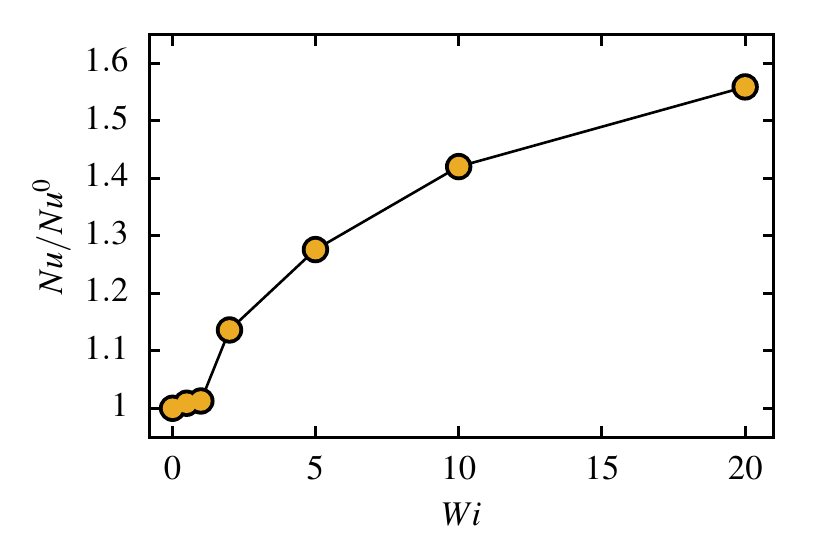}
\caption{}
\label{fig. 2c}
\end{subfigure}
\caption{Effect of elastic turbulence on the global flow quantities: (a) Root-mean-square of velocity fluctuations, $U_{rms}$, normalized by the mean flow velocity, (b) dependence of the global friction factor, $f$, normalized by $f^0$, and (c) dependence of the global heat transfer, $Nu$, normalized by $Nu^0$ as a function of $Wi$. where $f^0$ is the friction factor for the laminar flow.
}
\label{fig: 2}
\end{center}
\end{figure}
\begin{figure}[!ht]
\begin{center}
\includegraphics[width=0.95\linewidth]{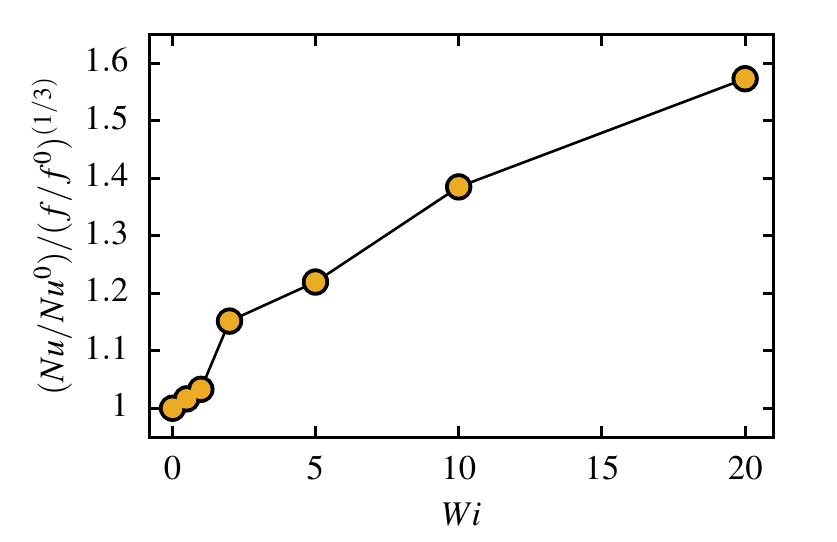}
\caption{Global Heat transfer performance in the serpentine channel for different $Wi$.}
\label{fig: 3}
\end{center}
\end{figure}
 \begin{figure*}[!ht]
\begin{center}
\begin{subfigure}[b]{0.45\textwidth}
\includegraphics[width=\textwidth]{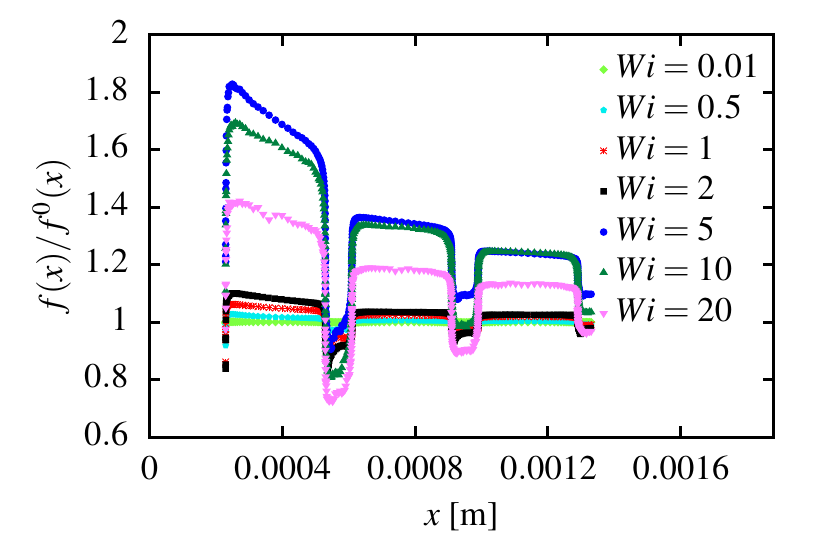}
\caption{}
\label{fig: 4a}
\end{subfigure}
\begin{subfigure}[b]{0.45\textwidth}
\includegraphics[width=\textwidth]{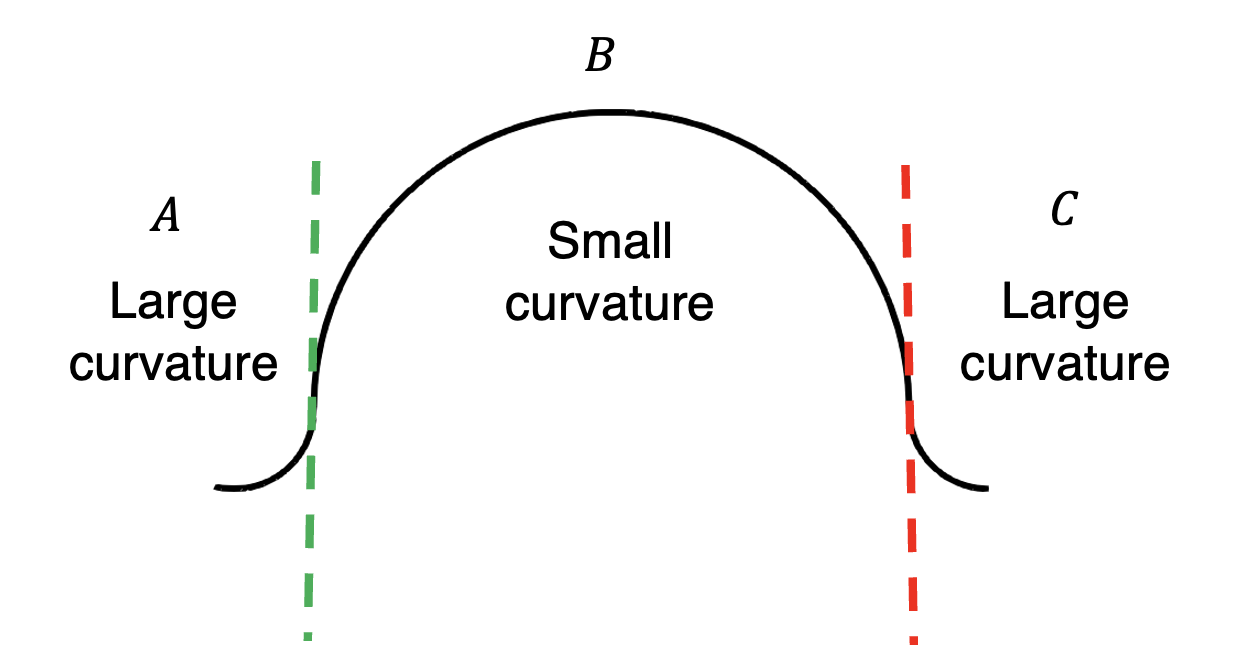}
\caption{}
\label{fig: 4a0}
\end{subfigure}

\begin{subfigure}[b]{0.45\textwidth}
\includegraphics[width=\textwidth]{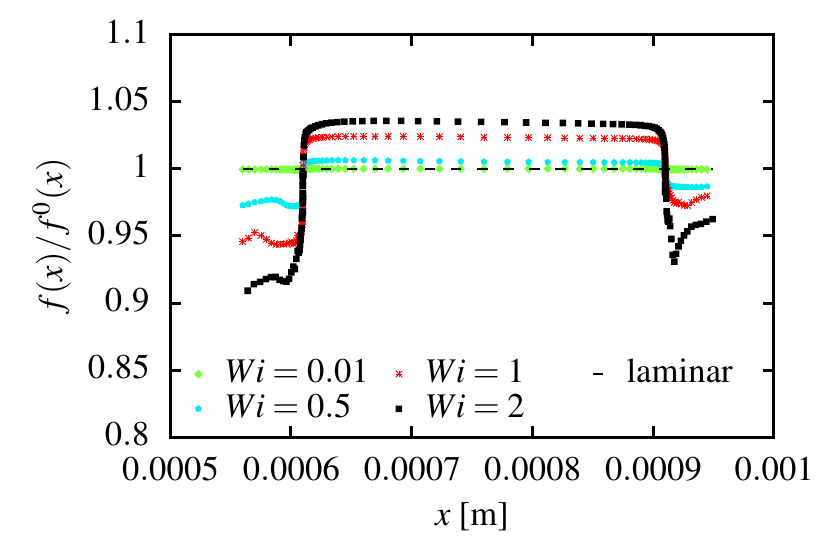}
\caption{}
\label{fig: 4b}
\end{subfigure}
\begin{subfigure}[b]{0.45\textwidth}
\includegraphics[width=\textwidth]{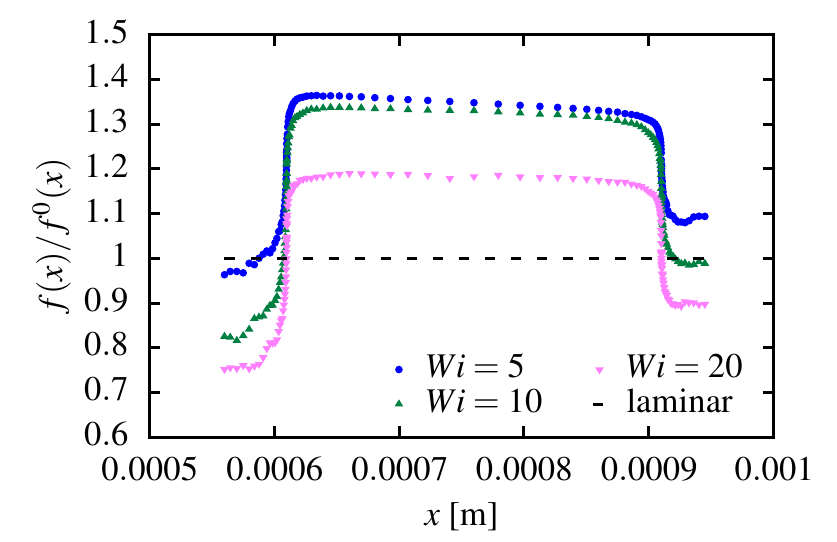}
\caption{}
\label{fig: 4c}
\end{subfigure}
\caption{Effect of elastic turbulence on the local friction factor: dependence of the local friction factor, $f(x)$, on $x$ { (a) along the top wall of the full channel, and (c, d) along one turn of the top wall}, as a function of $Wi$. Panel (b) represents the schematic {of one turn of the top wall} involving different curvatures (for illustration purposes). The values of $f(x)$ are normalized by $f^0(x)$, where $f^0(x)$ is the local friction factor for the laminar flow in the serpentine channel.}
\label{fig: 4}
\end{center}
\end{figure*}
\begin{figure}[!ht]
\begin{center}
\begin{subfigure}[b]{0.45\textwidth}
\includegraphics[width=\textwidth]{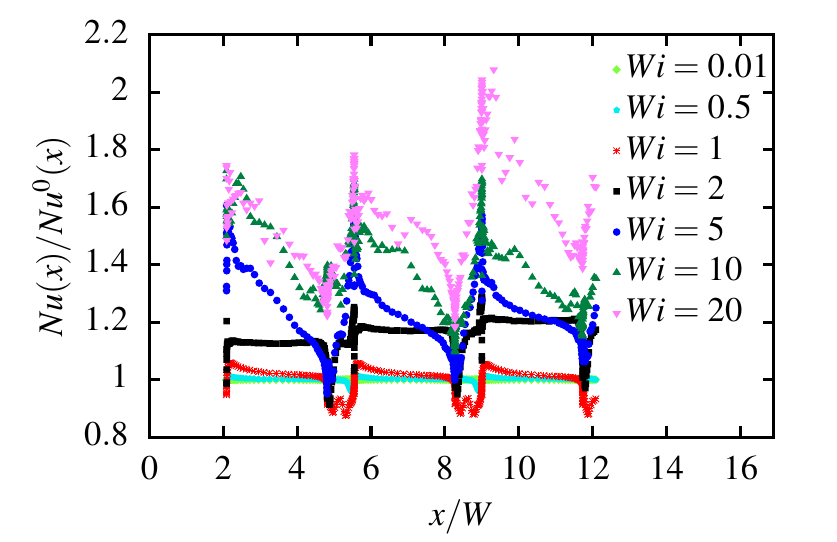}
\caption{}
\label{fig: 5a}
\end{subfigure}

\begin{subfigure}[b]{0.45\textwidth}
\includegraphics[width=\textwidth]{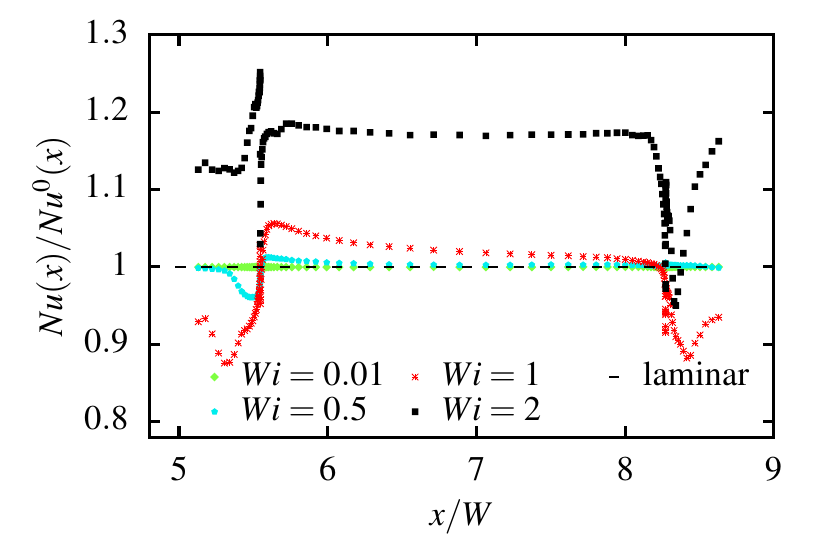}
\caption{}
\label{fig: 5b}
\end{subfigure}

\begin{subfigure}[b]{0.45\textwidth}
\includegraphics[width=\textwidth]{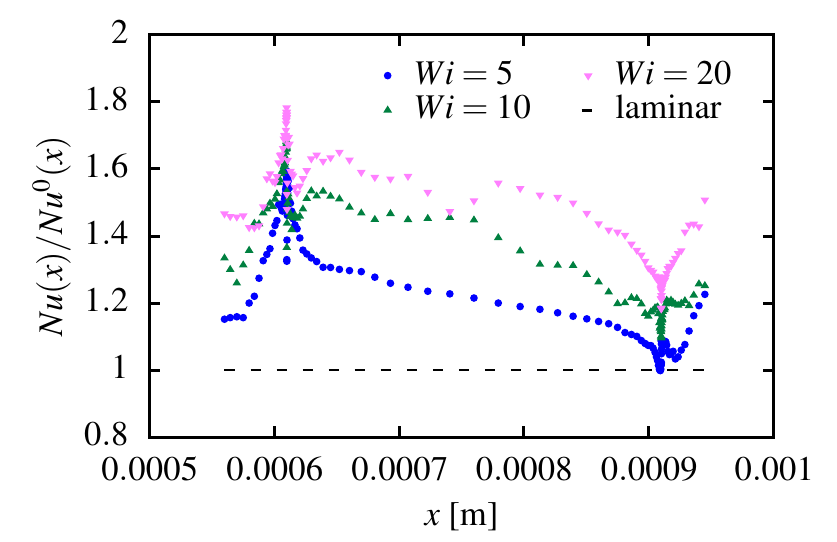}
\caption{}
\label{fig: 5c}
\end{subfigure}
\caption{Local heat transfer performance in the serpentine channel for different $Wi$: dependence of local Nusselt number, $Nu(x)$, on $x$ (a) along the full channel and (b, c) along one turn of the channel, as a function of $Wi$. The values of $Nu(x)$ are normalized by $Nu^0(x)$, where $Nu^0(x)$ is the local Nusselt estimation for the laminar flow in the serpentine channel.}
\label{fig: 5}
\end{center}
\end{figure}
\subsection{Global estimations of pressure drag and heat transfer}
Previous research has demonstrated that elastic turbulence can improve the transfer of mass, mixing effects, and overall heat transfer performance. In this study, we quantitatively examine the impact of elasticity on friction factor and heat transfer performance by varying only $Wi$. Our analysis focuses on global estimates of turbulence intensity ($U_{rms}/\left< U\right>$), average friction factor ($f=\sum_{x=1}^Nf(x)/N$), and average Nusselt number ($Nu=\sum_{x=1}^NNu(x)/N$), where $N=5,000$ is the number of point-like probes in the area being considered. All friction factors and Nusselt numbers presented are normalized by the respective values for a Newtonian laminar flow (benchmark case).
 
Figure \ref{fig. 2a} illustrates the normalized turbulence intensity ($U_{rms}/ U$). Three different flow regimes are observed: for $Wi< 3$, the turbulence intensity is infinitesimally small, indicating a laminar flow regime; for $3<Wi\leq 5$, a sharp increase is observed, indicating the onset of elastic turbulence, and $Wi$ is referred to as the critical Weissenberg number ($Wi_{crit}$); and finally, for $5<Wi\leq 20$, a continuous increase approaching an asymptotic value of $0.125$ is observed, indicating a fully developed elastic turbulence regime. The critical value of $Wi \sim 3-4$ is consistent with experimental observations in serpentine channel flow \cite{Groisman2004, Burghelea2004}.

Figures \ref{fig. 2b} and \ref{fig. 2c} show the averaged friction factor and Nusselt number measurements along the channel, respectively. Three distinct flow regimes are identified, consistent with the previous results. In the laminar flow regime i.e. $Wi< 3$, the friction factor ($f/f^0$) decreases by approximately $5\%$, and heat transfer ($Nu/Nu^0$) increases by $< 1\%$. \textcolor{black}{As the polymer stretching increases and $Wi$ approaches the critical range ($Wi=3-5$), a sharp increase in $f/f^0$ is recorded (up to $20\%$), while $Nu/Nu^0$ continues to increase (up to 28\%). We speculate that the repeated turns and changes in the direction of the micro-serpentine channel enhance the effect of vortices and flow instabilities, leading to a non-monotonic behavior of the friction factor as $Wi$ increases.} This non-monotonic behavior can also be influenced by the complex interplay between elastic and inertial forces in the flow and the channel geometry. However, in the fully developed elastic turbulence regime, a sharp linear drop (up to 20\%) in $f/f_0$ is observed, which is associated with augmented heat transfer, as $Nu/Nu^0$ increases by approximately $60\%$ compared to Newtonian laminar flow. This sharp drop in $f/f^0$ is likely due to the polymer drag reduction phenomenon that is typically observed in high $Re$ polymeric turbulent flows, also known as elastio-inertial turbulence. A recent experimental work \cite{Jha2020} reported a similar observation. However, we did not observe the saturation regime with increasing $Wi$ in the rectangular serpentine channel flow, as reported in Abed \textit{et al.} \cite{Abed2016}. This could be due to the narrow range of $Wi < 75$ used in the present study.
Overall, these results anticipate that $Wi \gg Wi_{crit}$ must be chosen to enhance the heat transfer performance in microchannels. 

The thermal performance, which evaluates the increased heat transfer based on the same area, is also considered. The following expression is then used:
\begin{equation}
(Nu/Nu^0)/(f/f^0)^{1/3},
\end{equation}
which is suggested by Han and Park \cite{HAN1988}. 
Figure \ref{fig: 3} shows the serpentine channel flow thermal performance with polymers. In the tested $Wi$ range, polymeric flows, irrespective of the flow regime, have higher thermal performance than Newtonian laminar flows as for $Wi>0$ the value of $(Nu/Nu^0)/(f/f^0)^{1/3}$ is always greater than one.
\subsection{Weissenberg number dependence} 
\label{Sec: local flow properties}
The dependence of the evolution of elastic turbulence in curved channels on both $Wi$ and the variation of channel curvature is significant. Therefore, the pressure loss along a channel is a local quantity. {To investigate the local behavior of pressure losses due to the evolution of elastic turbulence, we analyzed the dependence of the local friction factor, $f(x)$, on the $x$-coordinate along the top wall of the channel over a wide range of $Wi=0.01-20$, as shown in Fig. \ref{fig: 4a}.} All the presented results have been normalized by those obtained for Newtonian laminar flow. The local value of the friction factor remains close to the benchmark case (Newtonian laminar flow) for $Wi<1$, and increases monotonically with increasing $Wi$, reaching its maximum value at $Wi=5$. \textcolor{black}{Further increase of $Wi$ results in a sharp drop in $f(x)/f^0(x)$. This phenomenon is similar to the results of Jha and Steinberg~\cite{Jha2020} even though the threshold $Wi$ number is different. In Jha and Steinberg~\cite{Jha2020}, the authors termed the regime of reduced $f(x)/f_0(x)$ as the drag reduction regime}. The overall trend is similar to that observed in global estimates of $f(x)/f^0(x)$. To highlight the differences between small and large ranges of $Wi$, separate plots are presented in Figs. \ref{fig: 4b} and \ref{fig: 4c} for $Wi<5$ and $Wi \geq 5$, respectively. For clarity, {only one turn of the top channel wall is considered.} The results clearly demonstrate that the friction factor is a non-linear function of $Wi$.

{To understand the impact of the top wall curvature on friction factor, we consider only one top wall turn and divide it into three distinct parts, namely $A$, $B$, and $C$, based on their curvature, as shown in Fig. \ref{fig: 4a0}. When $Re$ approaches zero, the pressure gradient forces are proportional to the non-linear feedback of the polymer conformation tensor, $\nabla \cdot \bm{\tau}$, as shown in Eq. (\ref{Eq: 2}).} This simplification helps us comprehend the non-linear behavior of $f(x)$ along the {top channel wall. In parts $A$ and $C$, where the top wall curvature is large, polymers close to the top wall get highly stretched, leading to an increase in $\nabla \cdot \bm{\tau}$ and $\nabla p$. As the top wall curvature descends, i.e., in part $B$, polymers close to the top wall start to relax, eventually leading to a linear decrease in $\nabla \cdot \bm{\tau}$ and $\nabla p$.} 
{This behavior is primarily attributed to the variation of the boundary layer thickness along the top wall of the channel, which alternates between inner wall (parts $A$ and $C$) and outer wall (part $B$), due to the centrifugal force~\cite{Baughn1987}.} 

{As the fluid flows through the curved channel, the centrifugal force acts outward, leading to a pressure gradient and a change in velocity distribution across the channel cross-section. This variation in velocity affects the development of the boundary layer, with the boundary layer becoming thinner on the outer wall compared to a straight channel flow. As a result, the thinner boundary layer and high shear stress on the outer wall tend to result in an increased friction factor. Conversely, a thicker boundary layer can contribute to smaller shear stresses and a smaller friction factor. This behavior is clearly observed in Figs. \ref{fig: 4b} - \ref{fig: 4c} along the top wall.} One can see that this pattern repeats itself due to the periodicity of the channel structure. 

We also conducted an investigation of heat transfer performance along a channel, examining the local estimation of the Nusselt number, $Nu(x)$, at different points along the serpentine channel and at only one turn of the channel, as shown in Fig. \ref{fig: 5}. Our analysis of friction factors showed that, regardless of the $Wi$ value considered, there was a significant enhancement in the Nusselt number compared to Newtonian laminar flows. This finding is consistent with the overall heat transfer performance. Furthermore, we observed that the difference in Nusselt number between adjacent sections became more significant from part $A$ to part $C$ (see Figs. \ref{fig: 5b} and \ref{fig: 5c}), which is strictly dependent on the curvature of the channel. We found that the regions with significantly enhanced heat transfer correspond to locations where polymer stretching is intense, which is also where the maximum values of $f(x)$ occur and vice versa. {Once again, this behavior is attributed to the variation of the boundary layer thickness along the inner (thick boundary layer) and outer (thin boundary layer) walls of the channel due to the centrifugal force.}
\section{CONCLUSIONS}
\label{Sec: 5}
{In this study, we have conducted a systematic investigation of the friction factor and heat transfer characteristics of a dilute polymer solution flowing through a 2D serpentine channel at a low Reynolds number of $Re=10^{-3}$ and a high Weissenberg number ranging from $Wi=0.01$ to $Wi=20$. Our results provide important insights into the impact of polymer-induced elastic turbulence on heat transfer in microfluidic systems.}

{At small values of $Wi=0.003, 0.01$, the polymers does not influence the flow behavior and stays in coiled state, and have shown by means of local and averaged values of friction factor and the Nusselt number. For $Wi \leq 3$ the friction factor was found to be lower than that of laminar flow, and a slight increase in the Nusselt number is observed (upto 12\%) indicating that elastic polymers can alter heat transfer and fluid flow properties even in the absence of turbulence. These findings have direct implications for microfluidics, offering enhanced flow control and manipulation possibilities in microfluidic devices.}

{For $Wi>3$, we observed a monotonic increase in heat transfer performance without a corresponding increase in the friction factor. The maximum friction factor was observed at $Wi=5$. Our analysis of heat transfer performance revealed a consistent monotonic increase with respect to $Wi$, suggesting that utilizing $Wi \gg Wi_{crit}$ in microfluidic channels can optimize heat transfer performance without significant pressure losses. Additionally, the examination of local heat transfer enhancement provided insights into regions of high polymer stretching and the importance of curvature in relation to elastic turbulence.}

{While the concepts of elastic instability and elastic turbulence have been explored in the literature for many decades, their studies have primarily focused on the origin and mechanisms. Only a few numerical studies have investigated the detailed heat transfer performance, and solely from a global perspective, in microfluidic systems. Hence, our analysis serves as an initial step towards understanding local thermal performances in the elastic turbulence regime and emphasizes the need for future numerical studies to explore this area in further detail.}
{In future studies, we intend to investigate these aspects more comprehensively by incorporating the 3D nature of the geometry and conducting complementary experiments, which will further enhance our understanding of the thermal performances in the elastic turbulence regime.} 
\section*{ACKNOWLEDGMENT}
The authors greatly appreciate the Swedish National Infrastructure for Computing (SNIC) for providing the Computer time, partially funded by the Swedish Research Council through grant agreement no. 2018-05973.
\section*{Author declarations}
\subsection*{Conflict of interest}
The authors have no conflicts of interest to disclose.
\subsection*{Data availability}
The data that support the findings of this study are available from the corresponding author upon reasonable request.
\newpage

\end{document}